\documentclass[11pt]{article}
\usepackage{amsmath, amssymb, authblk, cite, fullpage, graphics, hyperref, setspace}
\usepackage[utf8]{inputenc}

\begin{document}

\title{Observable vacuum energy is finite in expanding space}

\author{Csaba Bal\'azs\footnote{email: csaba.balazs@monash.edu}}

\affil{\small School of Physics and Astronomy, Monash University, Melbourne, Victoria 3800 Australia}

\date{}

\maketitle

\begin{abstract}

In this work I reason that in expanding space only those quantum modes contribute to the measured vacuum energy that do not transcend the observable volume.  Since all quantised field modes have various observable consequences, when a gravitational horizon causally confines an observer to a finite volume quantised modes should be restricted to the observable patch to remain consistent with gravity.  

Within the observable patch of Friedmann-Lema{\^\i}tre-Robertson-Walker (FLRW) space the vacuum expectation value of the energy-momentum tensor can be expressed as a sum over discrete field modes.  Friedmann's first equation provides a straightforward ultraviolet cut-off allowing only a finite number of modes in the sum.  The finite volume acts as an infrared regulator and the calculation of the vacuum energy density is tractable without regularisation and renormalisation.  

To test the validity of this idea I quantise a scalar field on an FLRW background and calculate its vacuum energy density in the vacuum dominated, conformal, holographic limit.  In this limit I show that the quantum vacuum energy density scales with the square of the Hubble parameter, consistently with gravity.  In this example quantum vacuum expands space while the horizon of the expanding space limits the energy density of the vacuum to the observed value.

\end{abstract}

keywords: quantum field theory; expanding space; finite volume; vacuum energy

\section{Introduction} 

% expanding space > finite observable <rho>
In this brief note I posit that in expanding space-time the observable energy density of quantum vacuum is limited to a finite value of the order of the Hubble parameter squared.  I support this argument by showing that when restricted to the observable patch, the vacuum energy density of a quantised scalar field is proportional to the square of the Hubble parameter when it is calculated in the vacuum dominated, conformal, holographic limit.  

% observable modes/volume
These statements are based on limiting quantised modes of the scalar field to the observable patch of an expanding classical background space-time, and the subsequent calculation of the vacuum expectation value of the diagonal components of its energy-momentum tensor.  Restricting quantisation to the observable patch is motivated by the fact that all modes included in the field expansion lead to directly observable consequences.  One of them is the energy density of the vacuum state, but they contribute to other observables such as pressure, entropy, or temperature.  Including and quantising modes in the field expansion that do not fit within the gravitational horizon is inconsistent with observability dictated by gravity.

% expansion > horizon > finite observable volume > discrete modes only
In this spirit, the finiteness of the vacuum expectation value of the energy-momentum tensor components is the consequence of the presence of a cosmological horizon in expanding space and of Friedmann's first equation.  The horizon encloses an observer in a finite volume where quantum fields admit a discrete spectrum.  After identifying the expectation value of the time-time component of the energy momentum tensor with the observed average vacuum energy density, Friedmann's first equation restricts the energy of the sum of the modes within the volume to a finite value.  This implies a finite number of quantised modes within the observable volume, which yields a finite vacuum energy density.

% vacuum domination > zero curvature > scale/conformal invariance (V = 0) > no particle production
As a concrete demonstration of this generic argument, I quantise a spin zero field on the background of Friedmann-Lema{\^\i}tre-Robertson-Walker space.  For simplicity I assume that the energy density of the field is dominated by that of the vacuum.  In this case, for non-interacting and massless fields, the system is scale and conformally invariant.  Particle creation due to expansion is negligible, the vacuum state is time independent, that is the initial vacuum state is the adiabatic vacuum.  
% Another major simplification is that in vacuum dominated space the energy-momentum tensor is fully determined by the energy density.  

% simple limit: rho \& N saturated (holography) > E = Emin & N = A > rho
In the above scenario the observable vacuum energy density is determined by the number and energy distribution of a finite number of quantum modes within the horizon.  This distribution becomes very simple in the holographic limit in which the number of degrees of freedom, that is the number of quantum modes, is saturated.  The number of modes can only be maximised if the energy of each mode is minimal and they reside on the surface.  These conditions render the vacuum energy calculable.  The calculation yields a quantum vacuum energy density proportional to the square of the Hubble parameter, as expected from gravity.

\section{Classical field equations}

This section lays out the classical field theory framework underlying this work.  Most of this formulation is available in the literature \cite{Birrell:1982ix, 1989aqft.book.....F, 2007iqeg.book.....M, 2017lgrc.book.....Y}.  I, however, reproduce some of it to clearly establish the context, and expose details necessary to understand my results.

% action
For the purpose of illustrating quantisation within the observable patch, I consider a generic, real, spin zero field, \(\phi (x)\), and assume that this scalar field is coupled to gravity as described by the widely used action:
\begin{equation}
  \mathcal{S} = \int d^4 x \, \sqrt{-g} \, \left( 
    \frac{1}{16 \pi} \mathcal{R}
  - \frac{1}{2} \xi \mathcal{R} \phi^2 
  + \mathcal{L}(\phi) \right) .
\label{eq:action}
\end{equation}
Here $g$ is the determinant of the four dimensional space-time metric $g_{\mu\nu}(x)$, the corresponding Ricci scalar curvature is the trace of the Ricci curvature tensor, $\mathcal{R} = g_{\mu\nu} \mathcal{R}^{\mu\nu} (g_{\mu\nu})$, and $\xi$ is the coupling strength between the Ricci curvature and the scalar field.  
The Lagrangian density,
\begin{equation}
\mathcal{L}(\phi) = 
 - \frac{1}{2} g_{\mu \nu} \partial^{\mu }\phi \partial^{\nu}\phi
 - \mathcal{V}(\phi) ,
\label{eq:Lagrange}
\end{equation}
encapsulates the dynamics of the scalar field, having a potential $\mathcal{V}(\phi)$ which contains the mass term $m^2 \phi^2/2$.  
I set the bare cosmological constant to zero and will rely on the negative pressure of the quantum vacuum of the scalar field to expand space, as explained later.
Throughout this work I use Planck units with the reduced Planck constant, the speed of light in vacuum, and Newton's gravitational constant set to unity: 
$\hbar = c = G =1$.

% Klein-Fock-Gordon, Einstein
Variation of the action with respect to $\phi$ and $g_{\mu\nu}$ yields the equations of motion
\begin{equation}
  \left( 
  - g_{\mu\nu} \nabla^{\mu } \nabla^{\nu} 
  + \xi {\mathcal R} 
  \right) \phi 
  + \frac{d\mathcal{V}}{d\phi} = 0 ,
\label{eq:Klein-Fock-Gordon}
\end{equation}
and
\begin{equation}
  {\mathcal G}_{\mu\nu} = 8 \pi {\mathcal T}_{\mu\nu}(\phi) ,
\label{eq:Einstein}
\end{equation}
describing the coupled dynamics of the gravitational and scalar fields.  
In the Klein-Fock-Gordon equation $\nabla^{\mu }$ denotes the covariant derivative:
$g_{\mu\nu} \nabla^{\mu } \nabla^{\nu} \phi = \partial^\mu(\sqrt{-g} \; g_{\mu\nu} \partial^\nu \phi)/\sqrt{-g}$.  
On the left hand side of Einstein's equation the Einstein tensor is
\begin{equation}
  {\mathcal G}_{\mu\nu} = {\mathcal R}_{\mu\nu} - \frac{1}{2} g_{\mu\nu} {\mathcal R} .
\end{equation}
On its right hand side the energy-momentum tensor can be written as \cite{Faraoni:2000wk}
\begin{equation}
  {\mathcal T}_{\mu\nu}(\phi) = 
  \partial_{\mu} \phi \partial_{\nu}\phi 
  + g_{\mu\nu} \mathcal{L}(\phi)
  - \xi \nabla_{\mu} \nabla_{\nu} \phi^2
  + \xi g_{\mu\nu} \nabla_{\alpha} \nabla^{\alpha} \phi^2 
  + \xi {\mathcal G}_{\mu\nu} \phi^2
  .
\label{Tmunu(phi)}
\end{equation}

% FLRW
I assume that the energy distribution of the scalar field is spatially homogeneous and isotropic on large scales.  Consequently, the metric takes the Friedmann-Lema{\^\i}tre-Robertson-Walker (FLRW) form, 
\begin{equation}
  g_{\mu \nu} dx^\mu dx^\nu = 
  \textsf{a}^2
  (-d\eta^2 + ds_3^2) ,
\label{FLRW}
\end{equation}
where the time dependent scale factor \textsf{a} relates conformal and comoving time: $ \textsf{a} d \eta  = dt$.  
The square of the line element of the three dimensional space with constant Gaussian curvature $\kappa$ is
\begin{equation}
  ds_3^2 = \frac{dr^2}{1 - \kappa r^2} + r^2 \left( d\theta^2 + \sin^2\theta d\varphi^2 \right) .
  \label{ds_3^2=}
\end{equation}  

% Friedmann
With the FLRW metric the Ricci tensor simplifies to \cite{Habib:1999cs}
\begin{equation}
  \mathcal{R}_{tt} = -3 \left( \dot{\textsf{H}} + \textsf{H}^2 \right) , \\
  \label{eq:R_tt}
\end{equation}  
\begin{equation}
  \mathcal{R}_{ij} = \left( \dot{\textsf{H}} + 3 \textsf{H}^2 + 2 \frac{\kappa}{\textsf{a}^2} \right) g_{ij} .
  \label{eq:R_ij}
\end{equation} 
Here the Hubble parameter is the normalised expansion rate, $\textsf{H} = \dot{\textsf{a}}/\textsf{a}$, and the dot denotes derivative with respect to comoving time $t$.  The Hubble parameter can be also expressed in conformal time: 
$ h = \textsf{a}'/\textsf{a} = \textsf{a} \textsf{H}$,  
where the prime denotes derivative with respect to conformal time $\eta$.

% FLRW > EoM
The equations of motion are consequently reduced to \cite{2017lgrc.book.....Y}:
\begin{equation}
% written in conformal time (cf. Ydri.nb):
   \phi'' + 2 h \phi' - \Delta_3 \phi + 6 \xi {\mathcal R} \phi + \textsf{a}^2 \frac{d\mathcal{V}}{d\phi}
  = 0 ,
  \label{eq:KFG-FLRW}
\end{equation}
and
\begin{equation}
  \textsf{H}^2 = \frac{8\pi}{3}\rho - \frac{\kappa}{\textsf{a}^2},
\label{eq:Friedmann1}
\end{equation}
\begin{equation}
  \frac{\ddot{\textsf{a}}}{\textsf{a}} = -\frac{4\pi}{3}(\rho + 3 \textsf{P}) .
\label{eq:Friedmann2}
\end{equation}
In the Klein-Fock-Gordon equation $\Delta_3$ is the 3-dimensional space-like Laplace operator. 

% perfect fluid, continuity
When deriving Friedmann's equations, it is assumed that the ensemble average of the energy-momentum tensor is that of a perfect fluid,
\begin{equation}
  \langle {\mathcal T}_{\mu \nu} \rangle = (\rho + \textsf{P}) u_\mu u_\nu + g_{\mu \nu} \textsf{P} ,
\label{eq:PerfectFluid}
\end{equation}
where $\rho$ and \textsf{P} are the average observed energy density and pressure of the fluid, and 
$u_\mu = (1,0,0,0)$ is the comoving speed of the fluid.
Combining Friedmann's equations yields the continuity equation, 
\begin{equation}
  \dot{\rho} + 3 \textsf{H} (\rho + \textsf{P}) = 0 ,
  \label{eq:Continuity}
\end{equation}
expressing conservation of energy for the fluid.

% T00
Since this work focuses on vacuum energy, I assume that vacuum dominates the energy density with an equation of state $\rho = -\textsf{P}$.  
Due to the simple equation of state the energy-momentum tensor is fully determined by its $tt$ component which, for the FLRW metric, can be written as \cite{Habib:1999cs}:
\begin{equation}
  {\mathcal T}_{00} = 
% in comoving time
% \frac{1}{2} \dot{\phi}^2 + \mathcal{V}(\phi) - \frac{1}{2 \textsf{a}^2} \phi \Delta_3 \phi + 6 \xi \textsf{H} \phi \dot{\phi} + 3 \xi \left( \textsf{H}^2 + \frac{\kappa}{\textsf{a}^2} \right) \phi^2 
% in conformal time (cf. Habib.nb)
  \frac{1}{\textsf{a}^2} \left( \frac{1}{2} (\phi')^2 - \frac{1}{2} \phi \Delta_3 \phi + \textsf{a}^2 \mathcal{V}(\phi) + 3 \xi \left( h^2 + \kappa \right) \phi^2 + 6 \xi h \phi \phi' \right)
  .
  \label{eq:T_00}
\end{equation}
% Lambda_eff
The continuity equation shows that for vacuum domination the vacuum energy density is unchanged in time.  As well known, this enables the vacuum to act as an effective cosmological constant, 
\begin{equation}
  \Lambda_{\rm effective} = 8\pi \rho .
  \label{eq:CosmCons}
\end{equation}
Thus, the negative pressure of the vacuum can, in principle, expand FLRW space \cite{Zeldovich:1967gd, Zeldovich:1968ehl}.  

% dH/dt = kappa = 0
Vacuum dominated FLRW space expands exponentially in time.  This follows by eliminating the Hubble parameter from Friedmann's equations applied to the vacuum dominated case: 
\begin{equation}
  \textsf{a}(t) = \textsf{a}(t_0) \exp \left( \sqrt{ \frac{8 \pi}{3} \rho} \, t \right) .
  \label{eq:a(t)}
\end{equation}
Consequently, in vacuum dominated space the Hubble parameter is constant in comoving time, $\dot{\textsf{H}} = 0$, and the Gaussian curvature of the vacuum dominated FLRW space vanishes: $\kappa = 0$.
% conformal
With vanishing Gaussian curvature the FLRW metric is explicitly Lorentz invariant and conformally flat.  To preserve the conformal invariance of the action, that is to keep the calculation as simple as possible, I choose to conformally couple the scalar field to gravity by setting $\xi = 1/6$, and eventually set the potential $\mathcal{V}(\phi)$ to zero \cite{Polchinski:1987dy, Luty:2012ww, Nakayama:2013is}.  Since the zero point energy problem is typically discussed in the context of a free field, neglecting self interactions of the scalar field should not affect the main conclusion of this work.  Including the mass of the scalar field in this framework is left for later investigation.  

% horizon, finite volume
Space-like hypersurfaces of FLRW space with positive Gaussian curvature have closed geometry with finite circumferential radius and finite volume.  With negative or vanishing Gaussian curvature the space is open, extending to infinity, and its space-like volume is infinite \cite{Misner:1974qy}.  
Embedded in an expanding FLRW geometry however, regardless of the value of the Gaussian curvature, an observer has causal access only to a finite patch of the full spatial volume due to the presence of the gravitational horizon \cite{1956MNRAS.116..662R}.  Beyond this horizon physical processes are unobservable at a given time \cite{Melia:2007sd, Bikwa:2011bc, Melia:2012yx, Lewis:2012yk}.  
% Consequently, the observer is effectively enclosed in the finite region enclosed by the horizon.  

There are various choices of the horizon which might quantitatively affect the calculated value of observables, but do not affect the fact that these observables have to be calculated with the existence of the horizon taken into account \cite{Davis:2003ad}.
In a spatially flat FLRW space the Hubble radius, $R = 1/\textsf{H}$, coincides with the apparent horizon which is interpreted as the causal horizon of the observer \cite{Bak:1999hd}.  For this reason in this work I refer to the comoving spatial region within the Hubble radius, $V = 4 \pi R^3/3$, as the observable volume.

\section{Quantising the scalar field}

After establishing the classical field context, in this section I quantise the scalar field on the observable patch of the FLRW background.  To simplify expressions, I treat in detail the vacuum dominated conformal case and focus on the calculation of the time-time component of the energy momentum tensor.  The methodology proposed here, however, is independent of the special case and should be valid in general.  

% Dirac
During canonical quantisation the field and its conjugate momentum are prescribed commutation relations.  After the field is expanded in orthonormal modes,
\begin{equation}
 \phi (x) = \int \frac{d^3p}{(2\pi)^3}
 \left( 
 {\rm a}_{\pmb{p}} {\rm u}_{\pmb{p}}  (x) +
 {\rm a}_{\pmb{p}}^\dagger {\rm u}_{\pmb{p}}^*(x)
 \right) ,
 \label{eq:WrongExpansion}
\end{equation}
these commutators imply commutation relations on the coefficients of the linearly independent modes, ${\rm a}_{\pmb{p}}$, each labelled by its comoving three-momentum $\pmb{p}$.  
Expansion (\ref{eq:WrongExpansion}) is valid for open spaces, such as the FLRW space with zero or negative Gaussian curvature, or a static space with vanishing expansion rate.  For closed FLRW space, with finite spatial volume, a discrete form of this expansion is used.  

% Klein-Fock-Gordon
For vanishing Gaussian curvature the momentum dependence of the mode functions is harmonic: 
\begin{equation}
 {\rm u}_{\pmb{p}}(x) = 
 \frac{e^{\imath \pmb{p}\pmb{x}}}{\textsf{a}(\eta)} \chi_{\pmb{p}}(\eta) .
 \label{eq:u_p(x)}
\end{equation}
These mode functions satisfy the Klein-Fock-Gordon equation of motion implying
\begin{equation}
 \chi''_{\pmb{p}} + \epsilon_{\pmb{p}}^2 \chi_{\pmb{p}} = 0,
 \label{eq:KFG4chi}
\end{equation}
with
\begin{equation}
  \epsilon_{\pmb{p}}^2 = 
  \pmb{p}^2 
  + \textsf{a}^2 \frac{1}{2 \phi} \frac{d\mathcal{V}}{d\phi} 
  + \left( \xi - \frac{1}{6} \right) {\mathcal R} .
\label{eq:onshell}
\end{equation}

% harmonic modes
Equations (\ref{eq:KFG4chi}) and (\ref{eq:onshell}) show that, for the conformal case with a vanishing potential, the time dependence of the mode functions is also harmonic,
\begin{equation}
 \chi_p(\eta) = \frac{1}{\sqrt{2 \epsilon_{\pmb{p}}}} e^{-\imath \epsilon_{\pmb{p}} \eta} ,
\label{eq:chi_p(eta)}
\end{equation}
with a time independent energy $\epsilon_{\pmb{p}}^2 = \pmb{p}^2$.  The phase and normalisation of the time dependent part of the mode functions follows from the Wornskian condition: $\chi_p \chi_p'^* - \chi_p' \chi_p^* = \imath$.
The simple harmonic spatial and temporal dependence is not surprising, since a scalar field coupled to gravity via $\xi \mathcal{R} \phi^2$ in FLRW space is equivalent with a scalar field over Minkowski space with $\textsf{a}(\eta)$ dependent mass and couplings \cite{Hochberg:1994ty}. 

% particle creation
It is well known that on a time dependent background space positive and negative frequency field modes defined at a given time will later mix \cite{Parker:1968mv, Grib:1969ruc, Zeldovich:1970si, Hawking:1975vcx, Birrell:1979ip, Grishchuk:1993ds, deGarciaMaia:1994tsf}.  The mixing of equal time creation and annihilation operators leads to a time dependent number operator.  This effect implies a changing particle number, creation of particles, during time evolution of the quantum system and renders the definition of vacuum state time dependent.
In more technical terms, creation and annihilation operators at a later time become linear combinations of creation and annihilation operators at an earlier time, that is they become connected by a time-dependent Bogoliubov transformation \cite{1958NCim....7..843V, 1958NCim....7..794B}.

% no particle creation
In isotropically and smoothly extending space however, such as the vacuum dominated FLRW space, there is no particle production of massless particles satisfying a conformal field equation \cite{Parker:1969au, Parker:1971pt}.  In the conformally symmetric case the mode equation admits harmonic solutions and the Bogoliubov coefficients are time independent.  This means that the adiabatic number basis coincides with the initial basis spanned by ${\rm a}_{\pmb{p}}$, that is the initial vacuum remains the adiabatic vacuum during expansion \cite{Parker:1969au, Habib:1999cs}.  Consequently, the vacuum can be defined time independently as the state annihilated by all ${\rm a}_{\pmb{p}}$: ${\rm a}_{\pmb{p}} | 0 \rangle = 0$ for all $\pmb{p}$. 

% rho
Substituting the field expansion (\ref{eq:WrongExpansion}) into the expression of the energy-momentum tensor (\ref{eq:T_00}), in the conformal and massless case, leads to an undefined average vacuum energy density of the form \cite{Habib:1999cs, Kaya:2011yu}
\begin{equation}
  \rho = 
  \textsf{a}^4 \langle 0 | {\mathcal T}_{00} | 0 \rangle = 
  \int %_0^\Lambda 
  \frac{d^3p}{(2 \pi)^3} \frac{1}{2} \epsilon_{\pmb{p}} .
% \simeq \frac{\Lambda^4}{16 \pi^2} ,
\label{eq:rho=int}
\end{equation}
Here, I identified the time independent part of the vacuum expectation value of the time-time component of the energy momentum tensor (\ref{eq:T_00}) with the average comoving vacuum energy density $\rho$.
The result in equation (\ref{eq:rho=int}) coincides with the one calculated in static ($\textsf{a} = 1$) and spatially flat FLRW (Minkowski) space, as expected \cite{Hochberg:1994ty}.

% CC pb
Without a cut-off the momentum integral in equation (\ref{eq:rho=int}) is divergent at its upper limit.  It is generally expected that gravity cuts off the ultraviolet end of this integral, but it is unclear how this might happen.  If, for example, the cut-off scale is chosen to be the Planck scale then equation (\ref{eq:rho=int}) yields an energy density over 120 orders of magnitude higher than the observed average cosmological energy density.  This mismatch is known as the zero point energy problem, which is an essential part of the vacuum energy and cosmological constant problems (for some reviews see \cite{Weinberg:1988cp, Carroll:1991mt, Peebles:2002gy, Padmanabhan:2002ji, Martin:2012bt, Sola:2013gha, Burgess:2013ara, Padilla:2015aaa, Bull:2015stt, Bass:2015yaa, Peracaula:2022vpx}).

% observability of modes
According to equation (\ref{eq:rho=int}), the quantum vacuum carries the energy of an infinite number of modes, highlighting the fact that the presence of each mode in the field expansion leads to observable consequences.  
The integral in equation (\ref{eq:rho=int}) diverges because it is over an infinite range of continuous momenta which, in the quantum context, corresponds to infinite volume.  
In expanding space, however, the observable volume is finite.  
Consequently, when calculating observables only those modes should be included in the expansion that do not transcend the observable volume.
Populating unobserved regions of space with observable quantum modes is inconsistent with the observability limit set by gravity.  

% entanglement?
Quantum modes could causally connect an observer beyond the horizon to another within via entanglement \cite{Hawking:1976ra}.  This mode of information transfer, however, quickly degrades with an increasing expansion rate of space \cite{Bak:2019zsk}.  
Since vacuum dominated space expands exponentially, I assume a sharp cut-off of all observable modes at the horizon.  Operationally, I quantise the field restricted to the observable patch, rather than allowing it to propagate in the full volume.  This is a simple approximation which could certainly be refined.

% Matsubara
In finite observable volume only a countable number of Fourier modes fit, and the field admits a discrete expansion \cite{Matsubara:1955ws}:
\begin{equation}
\phi (\pmb{x}) = \sum _{i=1}^N \frac{1}{\sqrt{2V \epsilon_{\pmb{p}_i}}} 
  \left( 
  a_{\pmb{p}_i} {\rm u}_{\pmb{p}_i}  (x) +
  a_{\pmb{p}_i}^\dagger {\rm u}_{\pmb{p}_i}^*(x)
  \right) .
\label{eq:RightExpansion}
\end{equation}
The three-momenta ${\pmb{p}_i}$, carrying an integer index, label the quantum modes which are allowed in the observable volume.
The operators creating and annihilating these modes satisfy the equal time commutation relations, 
\begin{equation}
\left[ a_{\pmb{p}_i}, a_{\pmb{p}_j}^{\dagger} \right] = \delta_{ij}, ~~~~~~ \left[a_{\pmb{p}_i},a_{\pmb{p}_j}\right]=\left[a_{\pmb{p}_i}^{\dagger}, a_{\pmb{p}_j}^{\dagger }\right]=0 ,
\end{equation}
valid for any quantum system with countable modes.
% Hubble V
The volume $V$ contained within the gravitational horizon is causally accessible to an observer.  In the rest of this work I identify $V$ with the Hubble volume, although this choice can be altered.  The total number of modes within the volume is given by $N$ which is unspecified at this stage.

% countable vacuum DoFs
Crucially, with the discrete field expansion, the average comoving vacuum energy density within the observable volume becomes
\begin{equation}
  \rho = 
  \textsf{a}^4 
  \langle 0 | {\mathcal T}_{00} | 0 \rangle = 
  \frac{1}{V} \sum_{i=1}^N  
% \left(n_i+\frac{1}{2}\right) \epsilon_{\pmb{p}_i} .
  \frac{1}{2} \epsilon_{\pmb{p}_i} .
  \label{eq:<0|T_00|0>=sum}
\end{equation}
The discrete field expansion qualitatively transforms the mathematical form of the vacuum energy density.  It is no longer an integral to be cut off by an upper limit, rather a sum over discrete energy values.  

% mode filtering: volume is regulator
The vacuum expectation value of the energy density on the left hand side of (\ref{eq:<0|T_00|0>=sum}) is expected to match the average energy density in Friedmann's first equation (\ref{eq:Friedmann1}).  After making that identification, equations (\ref{eq:Friedmann1}) and (\ref{eq:<0|T_00|0>=sum}) require the energy of each mode to be less than the Hubble radius in natural units: $\epsilon_{\pmb{p}_i} \leq R$.  In other words, Friedmann's first equation acts as a straightforward ultraviolet regulator on the individual modes.  
The finiteness of the volume further demands that the energy of each mode be higher than the order of the inverse of the Hubble radius: $1/R \lesssim \epsilon_{\pmb{p}_i}$. Expressed differently, the finiteness of the volume acts as an infrared cut-off on the modes.  Consequently, the energy of all modes in the discrete sum in question has to be finite and within the $1/R \lesssim \epsilon_{\pmb{p}_i} \leq R$ window.  Under such conditions the sum in equation (\ref{eq:<0|T_00|0>=sum}) is finite provided its number of terms, $N$, is a finite integer.  

% vacuum structure
A finite number of observable quantised modes allowed in the vacuum, that is quantum vacuum having structure, is in sharp contrast to the notion of the ground state in infinite space where all modes are always present in the vacuum and contribute to its observable properties, such as its energy.  Nevertheless, in the presented framework, gravity is limiting the quantum vacuum energy via observability by restricting the number of degrees of freedom to a finite value inside the horizon \cite{Ford:2009vz}.  Since the gravitational horizon acts as an ultraviolet and infrared regulator, further regulation and renormalisation is unnecessary in this framework, which has far reaching consequences both for quantum field theory and for the quantum theory of gravity.

\section{The holographic limit}

Having found a regular, finite expression for the vacuum energy density, in this section I evaluate it in the holographic limit in which the number of modes over the volume is maximised.  In this case the number and energy distribution of the modes becomes trivial and the vacuum energy density can easily be calculated.

% holograf
In general, finding $N$ and a set of $\epsilon_{\pmb{p}_i}$ that determines the number and energy distribution of modes within the volume is non-trivial.  But in the holographic limit this task is relatively easy.  Following reference \cite{Balazs:2019cnm}, I consider the case when the number of quantum modes, $N$, is saturated within the Hubble volume.  
Holography postulates that the maximal number of degrees of freedom within certain gravitational horizons equals the area surrounding the volume measured in Planck units \cite{'tHooft:1993gx, Banks:1993en, Susskind:1994vu, Strominger:1996sh, Fischler:1998st, Bousso:1999xy, Bousso:2002ju}.  In case of vacuum dominated, conformal FLRW space, gravitational holography fixes the maximal value of the degrees of freedom in terms of the Hubble radius \cite{Bousso:1999xy}:
\begin{equation}
  N_{\max} = 4\pi R^2 .
  \label{eq:N_max}
\end{equation}

% saturation
Since for such a holographic system the number of degrees of freedom is saturated, with a fixed energy density and volume in equation (\ref{eq:<0|T_00|0>=sum}) the energy of the individual modes, $\epsilon_{\pmb{p}_i}$, has to be as small as possible.  The minimal energy value of the field modes is easily calculated by observing that harmonic modes saturate Heisenberg's uncertainty relation $\Delta x \Delta p \geq 1/2$ \cite{1988AmJPh..56..318H}.  The maximal spatial variance in a spherical configuration, $\Delta x = 2 \pi R$, is achieved if the mode is localised near the surface \cite{Balazs:2019cnm}.  This is consistent with the holographic expectation that the degrees of freedom reside on the surface.  For a massless mode the uncertainty relation implies:
\begin{equation}
  \epsilon_{\min} = \frac{1}{4 \pi R} .
  \label{eq:eps_min=}
\end{equation}

% vacuum = Lambda 
Substituting $N = N_{\max}$ and $\epsilon_{\pmb{p}_i} = \epsilon_{\min}$ into equation (\ref{eq:<0|T_00|0>=sum}) yields:
\begin{equation}
  % \textsf{a}^4 
  \rho = \frac{3}{8 \pi} \textsf{H}^2 ,
  \label{eq:rho=3/8piH^2}
\end{equation}
where I also used the definition of the Hubble volume.  Notably, the calculation of the quantum vacuum energy density in the holographic scenario yields Friedmann's first equation, despite the fact that Friedmann's equations were only used to establish the finiteness of $N$ and $\epsilon_{\pmb{p}_i}$, but not their value.  Consequently, in this context, vacuum energy expands space while expanding space limits the energy of the vacuum to the observed value.  

% 10^121
Based on the above insight in a fictitious, vacuum dominated, conformal, holographic universe where the Hubble radius is the Plank length, $R = 1$, the number of maximal degrees of freedom is the Planck area, $N_{\max} = 4 \pi$, and the corresponding minimal energy is its inverse, $\epsilon_{\min} = 1/(4 \pi)$.  In such a universe the ultraviolet and infrared cut-offs are both the Planck scale, and the vacuum energy density is $\rho_{fictitious} = 3/(8 \pi)$.  If we confuse our universe with the fictitious one and apply the Planck scale as an ultraviolet cut-off to our universe, instead of the actual ultraviolet regulator (the present Hubble radius, $R_0 = 1/\textsf{H}_0$), we will arrive at a discrepancy of $\rho_{fictitious}/\rho_{observed} = 1/\textsf{H}_0^2$ when calculating the vacuum energy density.  The value of $1/\textsf{H}_0^2$ is $7.2 \times 10^{121}$ in Planck units.

\section{Conclusions}

% mode filtering abstract
In this work I argued that fields should be quantised on the observable patch of space to remain consistent with gravity.  In expanding space gravity renders the observable volume finite via a causal horizon.  Quantum modes that do not fit within the horizon cannot be observable and should not be included in the field expansion.  In finite volume the field admits a discrete spectrum and Friedmann's first equation, which sets an upper limit on the energy density within the volume, requires the spectrum to contain only a finite number of quantum modes.  Consequently, in expanding space the quantum vacuum energy is finite.  

% GRAVITY CUTS OFF THE NUMBER OF DOFS!
According to the above argument the horizon filters out unobservable quantum modes.  Gravity limits the number of degrees of freedom within the observable volume to a finite value, cutting off the energy density of quantum vacuum.  The energy and momenta of quantum modes are also regulated in the infrared by the finiteness of the observable volume in expanding space.  Notably, the infrared regulator, \textsf{H}, is the inverse of the ultraviolet regulator, $R$, suggesting self-duality.  Thus, in this scenario the vacuum energy density is rendered small by natural environmental regulators, the Hubble scale and its inverse, without introducing any fine-tuning.  

% vacuum structure
Intriguingly, a finite number of of observable modes within the horizon implies that the vacuum within the observable volume acquires structure.  While in infinite space vacuum contains all Fourier modes, in an expanding universe the vacuum may only contain a finite number of observable field modes with a particular number and energy distribution.  The potentially tractable structure of the vacuum in finite observable volume profoundly affects the fundamentals of quantum field theory and quantisation of gravity.

% connections
The presented framework has similarities to recently proposed solutions of the vacuum energy problem \cite{Bousso:2007kq, Ganesh:2019tfr, Park:2021vro, Firouzjahi:2022xxb}.  Especially, the idea of a finite number of quantised degrees of freedom attracted recent attention in the context of holography \cite{Dvali:2013eja, Bousso:2014uxa, Dvali:2015rea, Dvali:2019jjw, Dvali:2019ulr}.  In this work the same idea is applied to the cosmological vacuum energy, successfully generating an effective cosmological constant.  
%Thus, it is not necessary for the trace of the quantised energy-momentum tensor, that is the quantum vacuum contribution to the cosmological constant, to vanish \cite{Donoghue:2020hoh, Savvidy:2021ahq}.  Rather the opposite happens: When quantised by respecting the observability limit set by gravity, the the quantum vacuum contribution to the cosmological constant is consistent with observation.  
%
While the idea that quantisation should respect observability set by gravity is quite general, various details of the presented calculation could certainly be improved.  Including mass and self interactions, including fermions, or more precisely determining how the modes are cut off at the horizon are challenges left for future work. \\

{\bf Acknowledgements} ~~~ This research was funded by the Australian Research Council through Discovery Project grants DP180102209 and DP210101636.

\bibliographystyle{unsrt}
\bibliography{paper} 

\end{document}